# Acoustic metamaterial with negative modulus and a double negative structure.

B. Sharma,* and C. T. Sun[†]

School of Aeronautics and Astronautics, Purdue University, Armstrong Hall of Engineering,

West Lafayette, IN 47907, USA

## Abstract

An acoustic negative bulk modulus metamaterial based on the concept of expansion chambers is proposed. It is shown that addition of a neck region to an ordinary expansion chamber improves its transmission loss characteristics at low frequencies and the resulting structure displays a negative bulk modulus behavior. Additionally, membrane based metamaterials are analyzed. Using FEM, the negative density behavior of a membrane carrying a center mass and of a tensioned membrane array is analyzed and the inherent similarity of the two designs is discussed. Further, the modified expansion chamber is combined with an array of stretched membranes and the resulting structure is analyzed for double negative behavior.

*bsharma@purdue.edu, [†]sun@purdue.edu

# 1. Introduction

Metamaterials, broadly defined as artificial composite materials engineered to obtain unusual properties not readily available in nature, have been the subject of considerable interest over the past decade. Electromagnetic metamaterials have been extensively studied and structures displaying negative effective properties have been obtained [1-7]. Investigators have proposed structures displaying similar negative properties for acoustic and elastic wave propagation. For example, it has been shown that negative effective parameters can be obtained using an array of a solid core material of high density and a coating of elastically soft material [8-10], or by using a mass-in-mass lattice system [11-14]. A negative bulk modulus was obtained for elastic metamaterials by using bubble contained water spheres [15]. Similarly, an array of Helmholtz resonators [16-18] and an array of side holes in a cylindrical duct [19] was shown to produce negative effective modulus for acoustic systems.

More recently, investigators have experimentally demonstrated the existence of a high transmission loss band-gap using stretched membranes [20-23]. It was shown that a high transmission loss band can be obtained between the first two eigenmodes of a clamped, circular membrane with an attached center mass [20,21]. This was attributed to a negative density behavior of the structure. Using an array of similarly clamped, circular membranes, however not carrying a center mass, Lee et al. [22] obtained a low transmission band gap below the first membrane resonance frequency and this too has been attributed to a negative density behavior. Thus, two systems based upon stretched membranes have been shown to display negative effective density in two different frequency regimes.

In this paper, we first present a structure capable of realizing the negative effective modulus behavior. The structure is based on the commonly used acoustic muffler, modified to include a neck in order to obtain a local resonating behavior. Next, we analyze the membrane based metamaterials displaying negative density behavior. Using finite element methods, we examine both designs and explain their inherent similarity. Using the obtained transmission and reflection coefficients, the effective parameters are calculated and it is shown that both systems display negative effective density associated with the membrane pretension and system antiresonance. Finally, the negative modulus structure was combined with an array of stretched membranes and a low frequency pass band was obtained.

## 2. Negative Bulk Modulus Design

Expansion chambers are commonly used as a simple model of a muffler and are routinely used as noise control devices in simple automobile mufflers, gun silencers, and sound-absorbing plenum chambers used in architectural acoustics [24-28]. The simplest expansion chamber design consists of an enlarged section of a pipe inserted in a sound transmission duct as shown in Fig. 1(a). Though simple and easy to install, its performance is limited by its length and area expansion ratio. To overcome these issues and to obtain improved performance at lower frequencies, we study the effect of addition of a neck region to the expansion chamber as shown in Fig. 1(b). Analysis of the proposed structure shows the existence of a negative effective bulk modulus region, similar to designs proposed by other researchers [16,19,23,29].

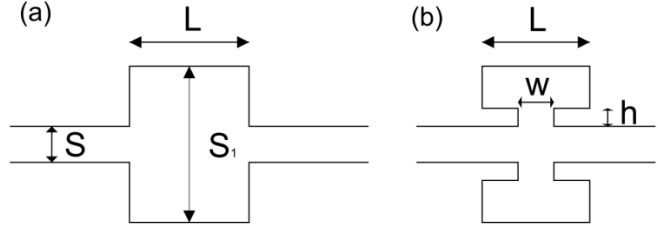

FIG. 1. (a) Traditional expansion chamber and (b) a modified expansion chamber with a neck region connecting the chamber with the main transmission duct.

The transmission characteristics for a 1D expansion chamber are readily available in the literature [24,26]. For an enlarged section of pipe of cross-sectional area $S_1$ and length L in a pipe of cross section S, under the assumption of continuity of pressure and volume velocity at the two junctions of the expanded pipe with the original pipe, the transmission coefficient, $T_\pi$, can be derived to be [24]:

$$T_\pi = 4/(4\cos^2(kL) + (S_1/S + S/S_1)^2 * \sin^2(kL)) \tag{1}$$

where k is the acoustic wavenumber. The transmission coefficient reaches a minimum value of

$$T_\pi = ((2S*S_1)/(S^2 + S_1^2))^2 \tag{2}$$

for $kL = \pi/2$, when the length of the filter section is a quarter wavelength.

Though these equations are valid only when the wavelength is large compared with the radius of the original pipe or with the dimensions of any filter section, they are useful for validation of a finite element model. An axisymmetric FEM model was constructed using the commercial finite element code ABAQUS. A duct of radius 12 mm with an expansion region of length 400 mm and area ratio 10 was modeled using quadratic acoustic elements. Density of air was taken as 1.21 Kg/m$^3$, the bulk modulus as 142 KPa and it was considered to be inviscid. The duct radius was chosen such that only the plane wave mode can propagate in the frequency range of interest [30]. An acoustic pressure wave of unit amplitude was sent in through the upstream end of the duct, while the downstream end was anechoically terminated by using the characteristic impedance of air. A linear perturbation, steady state analysis was performed and the output pressures were obtained in the frequency domain. Transmission characteristics were then obtained using the four microphone transfer matrix method as described in ASTM E2611-09 [31,32]. The obtained transmission coefficient was compared with the transmission coefficient obtained using Eq.1. The simulation results matched perfectly with the theoretical results in the plane wave transmission region.

Keeping the duct radius as 12 mm, the chamber length, L, was reduced to 30 mm and the radius of the expansion chamber was taken as 58 mm. Considering a neck of width (w) 2 mm, the neck height (h) was progressively decreased from 9 mm to 1 mm and then, keeping the neck height at 6 mm, the neck width was varied from 30 mm (no neck) to 2 mm. Figure 2(a) shows the effect of variation of neck height while Fig. 2(b) shows the effect of variation of neck width. It can be seen that as w decreases, the transmission loss at low frequencies increases, with transmission peaks occurring at 923 Hz and 460 Hz for neck width of 10 mm and 2 mm, respectively. For variation of neck height, it can be seen that as h is increased, the transmission loss peak shifts to a lower frequency.

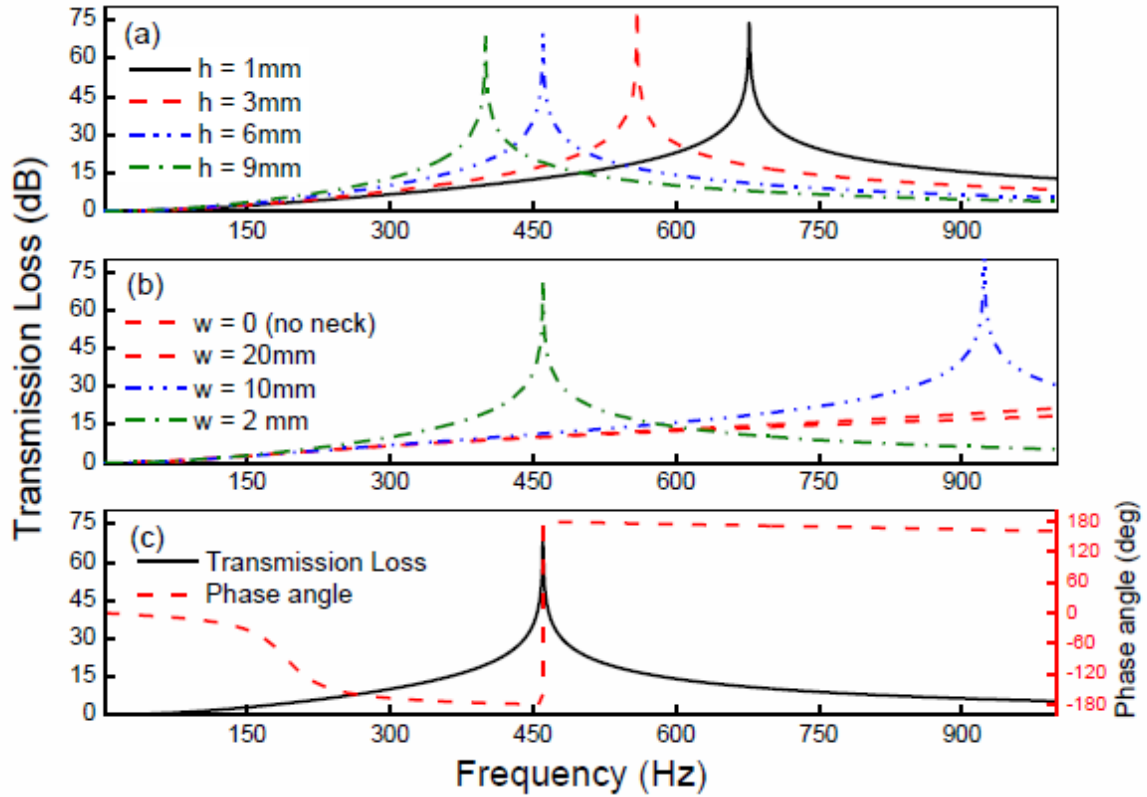

FIG. 2. (a)Variation of transmission loss with variation of neck height for a neck of constant width of 2 mm. (b)Variation of transmission loss with variation of neck width for a neck of constant height of 6 mm. (c) Transmission loss and phase change (right hand axis) as measure inside the chamber with a neck of height 6 mm and width 2 mm.

This high transmission loss peak can be attributed to the resonance characteristics of the expansion chamber as can be seen by the phase angle variation as measured in the expansion chamber and shown in Fig. 2(c). This resonance phenomenon leads to an accumulation of energy over many cycles and is significant to maintain the sequence of displacement near resonance even when the excitation field changes sign [16]. As the incident pressure field is swept through the resonance, the instantaneous displacement of the mass center in the unit changes from in-phase to out-of-phase with the driving field and we would expect the material to show a negative response. The effective properties for the above expansion chamber geometry, with neck width 2 mm and height 6 mm, were obtained by using an inverse technique [33-35]. Using the reflection and transmission coefficients obtained by the transfer matrix method, the effective mass density and bulk modulus were obtained under the conditions that the real part of the impedance ratio and the imaginary part of the sound speed are always positive. The real parts of the effective mass density $\rho_{eff}$ and bulk modulus $K_{eff}$ are as shown in Fig. 3. As expected, the effective properties are frequency dependant and the real part of the bulk modulus becomes negative in the neighborhood of the high transmission loss region and stays negative till about 1600 Hz.

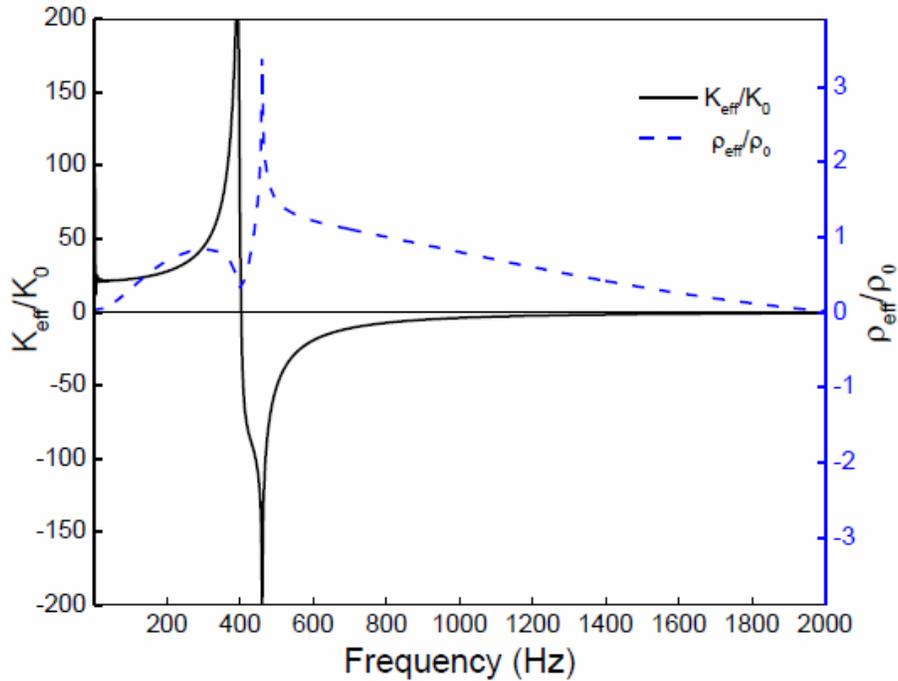

FIG. 3. Effective modulus (left axis, solid curve) and effective density (right axis, dotted curve) of a modified expansion chamber with a neck of length 6 mm and width 2 mm.

To obtain a forbidden band region, an array with six expansion chambers was modeled and the transmission coefficient is shown in Fig. 4. The pressure transmission shows the local peaks and dips equal to the number of chambers in the pass band between 0 Hz to 400 Hz, and a wide bandgap region is seen between 400 Hz to 1600 Hz. Thus, a wide, low frequency bandgap is achieved by using the negative effective modulus behavior of the modified expansion chambers.

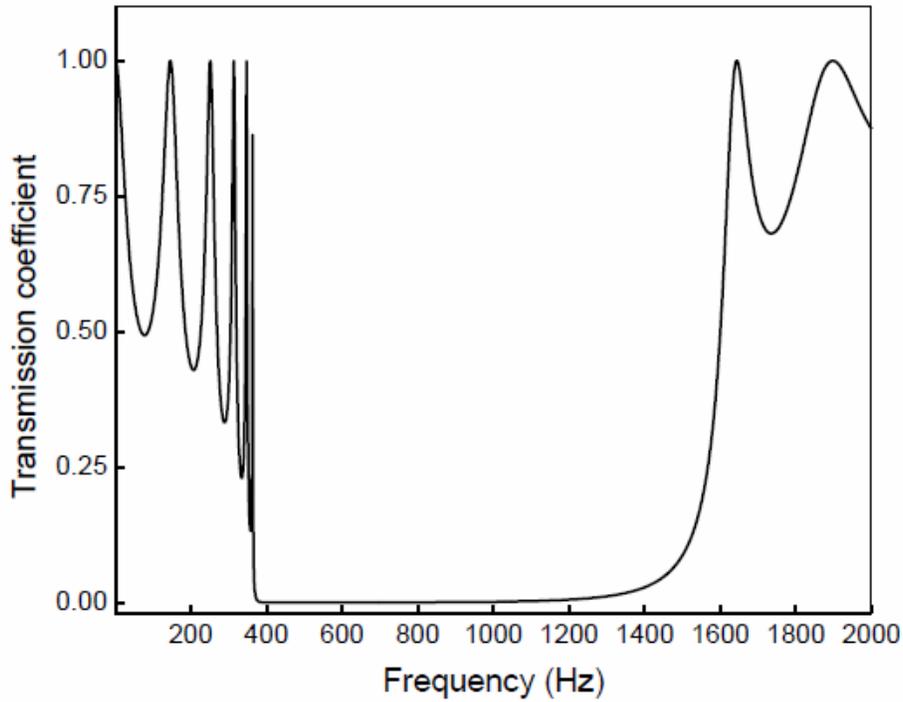

FIG. 4. Transmission coefficient of a modified expansion chamber array.

## 3. Membrane based metamaterials

It has been recently shown that near total reflection of acoustic waves can be obtained at a frequency between the first two eigenmodes of a clamped, circular membrane loaded with a finite mass at its center [20,21]. At frequencies around the total reflection frequency the in-plane average displacement normal to the membrane is minimum, leading to very small far-field transmission. Using volume averaged stress and accelerations over the whole membrane structure, a negative effective density was shown to exist. Similarly, by using an array of clamped, circular, stretched membranes, however without a center mass, Lee et al. [22] obtained a cut-on frequency below which all waves were demonstrated to be attenuated. Based on an approximate, lumped mass theory, this behavior was attributed to the system displaying negative effective density below the first membrane eigenmode. Thus, though both systems are based on stretched membranes, negative effective density was obtained in two different frequency regimes. To understand this further, fully coupled acoustic-structural models of the two systems were created using ABAQUS. Transmission characteristics and effective properties of the systems were obtained as described in the previous section.

A single membrane fixed in a rigid cylindrical impedance tube was modeled. An axisymmetric membrane of radius 12 mm, thickness 0.0762 mm and density 1200 Kg/m$^3$ was modeled using quadratic membrane elements. The elastic modulus and Poisson's ratio for the membrane material were taken as 6.9 GPa and 0.36, respectively. The membrane was prestressed by applying a radial stress of 0.65 MPa [36]. Quadratic acoustic elements were used to model the air inside the impedance tube and tie constraints were used to connect the membrane with the acoustic elements [37]. Again, the density of air was taken as 1.21 Kg/m$^3$, the bulk modulus as 142 KPa and it was considered to be inviscid. A steady state analysis, as described in the previous section, was performed and the output pressures were obtained in the frequency domain. Appropriate element sizes were taken so as to conform to the restriction of having a minimum of 6 elements per wavelength at the frequency of interest [37]. Convergence analysis was performed to confirm the accuracy of the data obtained. To fully understand the cut-on behavior, an array of 8 membranes was also modeled. The distance between the membranes was taken as 50 mm and pressure was measured after each membrane.

Along similar lines, a stretched membrane with a center mass was also modeled. The geometrical and material data were taken directly from [21]. The center mass was modeled using quadratic shell elements of the same thickness as the membrane. The density of the material used for the mass was chosen so as to match the total mass used for the experiments. The same procedure as explained above was used to conduct the simulation.

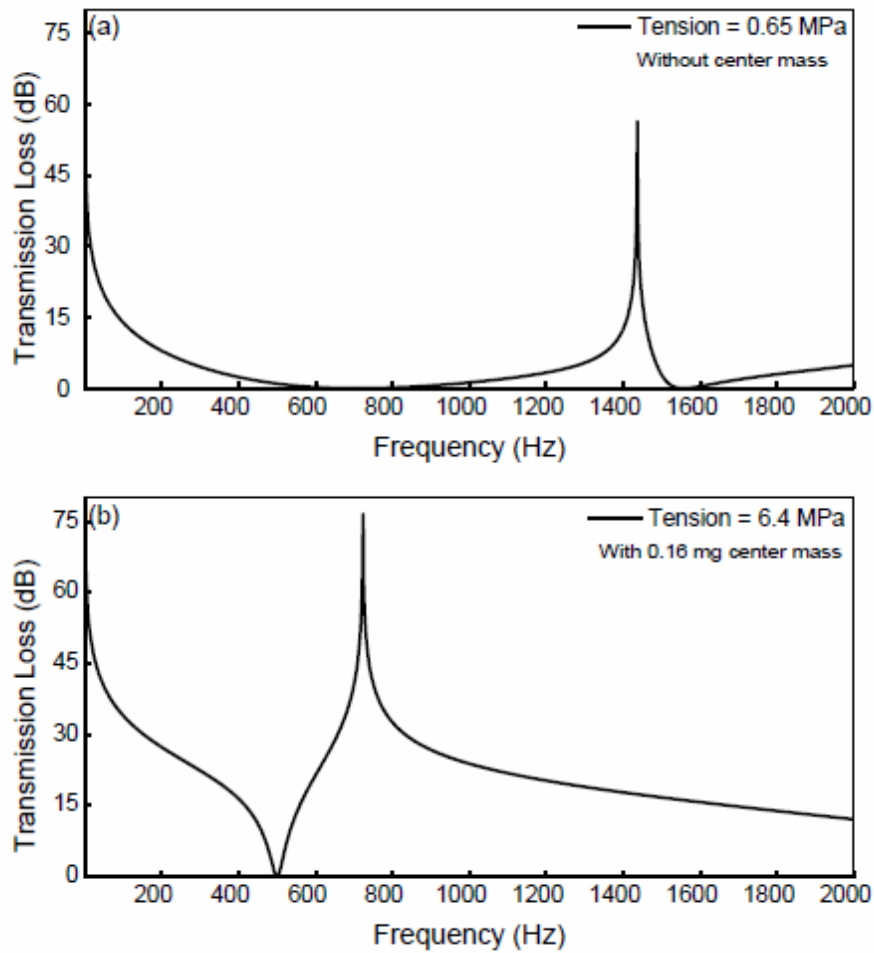

FIG. 5. (a) Transmission loss for a clamped circular membrane without a center mass. (b) Transmission loss for a clamped circular membrane with a 0.16 mg center mass.

Figure 5(a) and 5(b) show the transmission loss behavior for a clamped, circular membrane and a clamped circular membrane with a center mass of 0.16 gm, respectively. For both cases, high transmission loss is observed at low frequencies, with the value dropping as the frequency approaches the first eigenvalue. Note that the prestresses for the two cases, 0.65 MPa and 6.4 MPa, respectively, are different, and that the effect of addition of a center mass is to reduce the first eigenvalue to a lower frequency. The low frequency loss is due to the resistive force, provided by the stretched membrane, acting against the incident acoustic pressure. Thus, the pressure beyond the membrane is reduced by a finite step. For an array of stretched membranes, this reduced pressure acts as the incident pressure for the next membrane and so on until the pressure decays down to a negligible value. Figure 6 shows the pressure after the 1st, 3rd, 6th and 8th membranes and the spatial formation of a forbidden band gap region below the first resonance frequency can be clearly seen. Thus, a low frequency forbidden gap exists for both cases and the effect of a center mass is to reduce the cut-on frequency for a given membrane prestress.

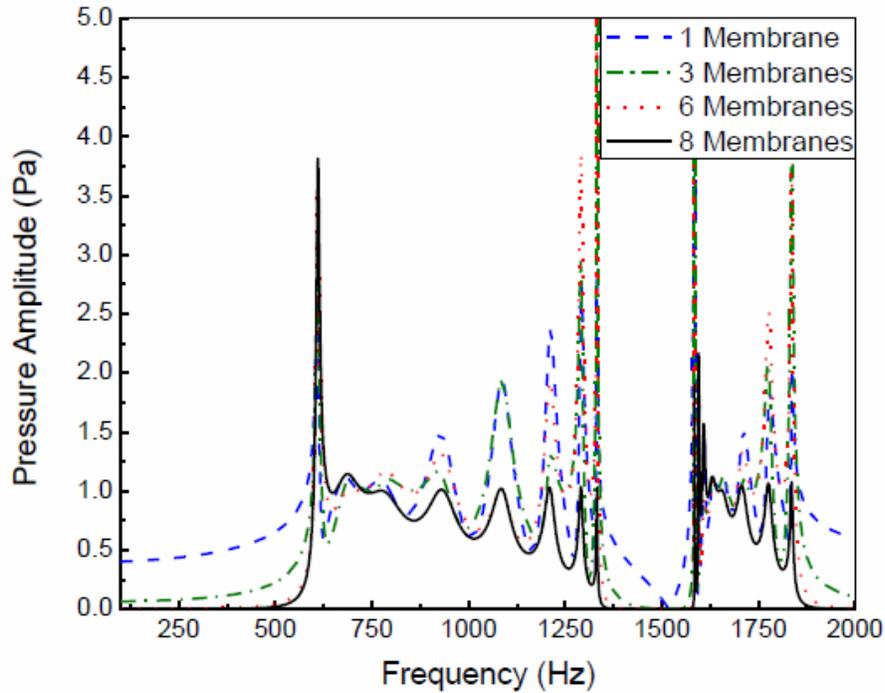

FIG. 6. Pressure amplitude with respect to frequency, plotted after 1 membrane (dashed line), 3 membranes (dash-dot line), 6 membrane (dotted line) and 8 membranes (solid line), respectively.

Between the first and second eigenvalues, in accordance with the reported experimental results [20,21], a high transmission loss region is seen for the mass loaded membrane. This has been shown to be caused due to the existence of an antiresonance frequency at which the average out of plane displacement is minimum, leading to near total reflection. For a clamped membrane without a center mass, at frequencies not reported in the experimental results, a similar high transmission peak can be seen close to the second resonance frequency. The high loss band in this case is narrower and at a higher frequency than for a mass loaded membrane. Thus, addition of a center mass reduces the anti-resonance frequency and broadens the high transmission loss region.

The effective properties were obtained as described previously. The variation of real part of the effective density of both systems with frequency is shown in Fig. 7. For both systems, negative effective density behavior can be seen in two different frequency regimes corresponding to the two mechanisms explained above. Thus, both systems are equivalent and the only effect of addition of a center mass is to tune the forbidden frequency gaps.

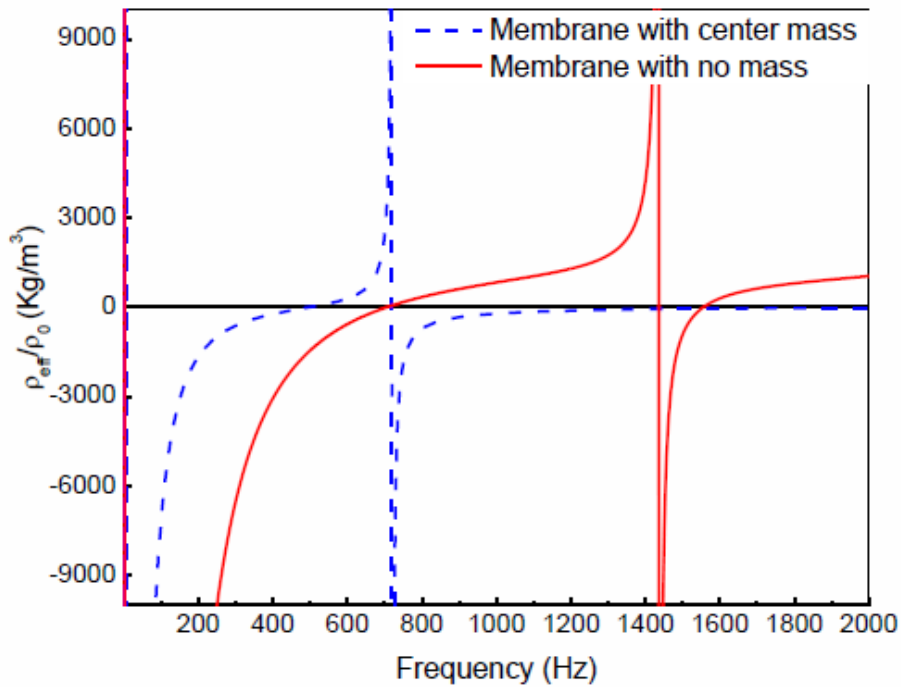

FIG. 7. Variation of the effective densities of membrane with mass (dashed line) and without mass (solid line) with respect to frequency. Only the real parts have been shown here for convenience.

## 4. Composite structure

Over the last few years, researchers have combined structures displaying negative effective density and negative effective modulus to obtain doubly negative structures [23,38.39]. Notably, [38] combined an array of stretched circular membranes in a tube with an array of side holes to obtain a pass band in the double negative region, while [23] replaced the side holes with axisymmetric shunts to obtain an acoustic transmission line metamaterial with negative refractive index. Based on this concept, the above presented modified expansion chamber was combined with a stretched membrane and the composite structure was analyzed.

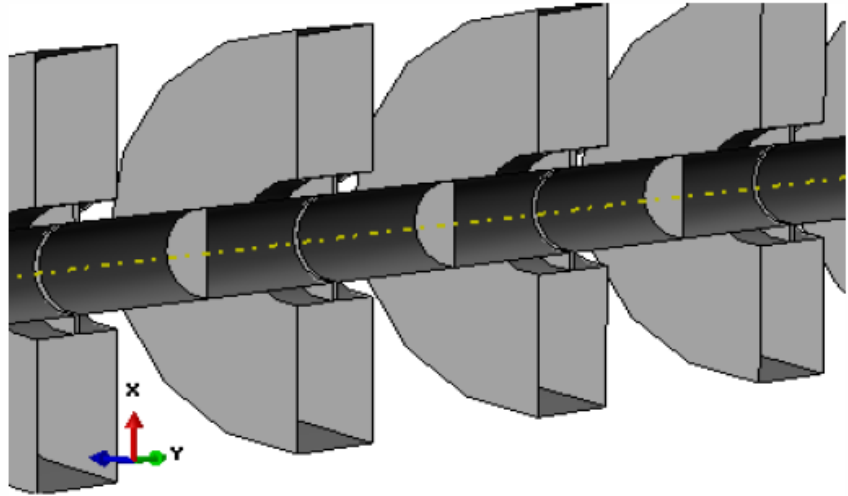

FIG. 8. Representative model of an array of the composite structure. An array of six unit cells was used.

Figure 8 shows a representative model of the combined structure. The membrane properties and dimensions were the same as used for the earlier case of stretched membranes without a center mass, while the modified expansion chamber dimensions were also taken to be the same as mentioned in the previous section. Based on the individual analysis, the effective density and modulus regions overlap in the 400 Hz to 705 Hz frequency band and in the 1450 Hz to 1600 Hz region. A unit cell was first modeled and the transmission behavior was obtained. The distance between the chamber opening and the membrane was taken sufficiently large in order to avoid the near field effects of the two systems affecting their behavior.

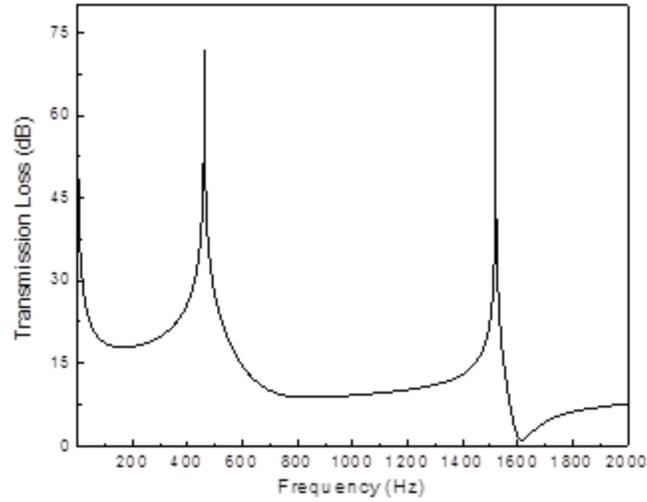

FIG. 9. Transmission loss curve of a composite structure unit cell.

Figure 9 shows the transmission loss obtained for the combined system. At low frequencies, the TL curve is seen to follow the TL curve of the stretched membrane. As the frequency increases and the modified expansion chamber starts affecting the transmission behavior, the TL curve slope changes and follows the behavior of the modified expansion chamber TL curve, peaking at the resonance frequency associated with the chamber, and then dipping to a value of 8 dB near 750 Hz. At higher frequencies, it follows the slope of the membrane TL curve, with a transmission peak at 1520 Hz. This peak is associated with the antiresonance behavior of the membrane and is shifted by about 80 Hz as compared to a membrane only system.

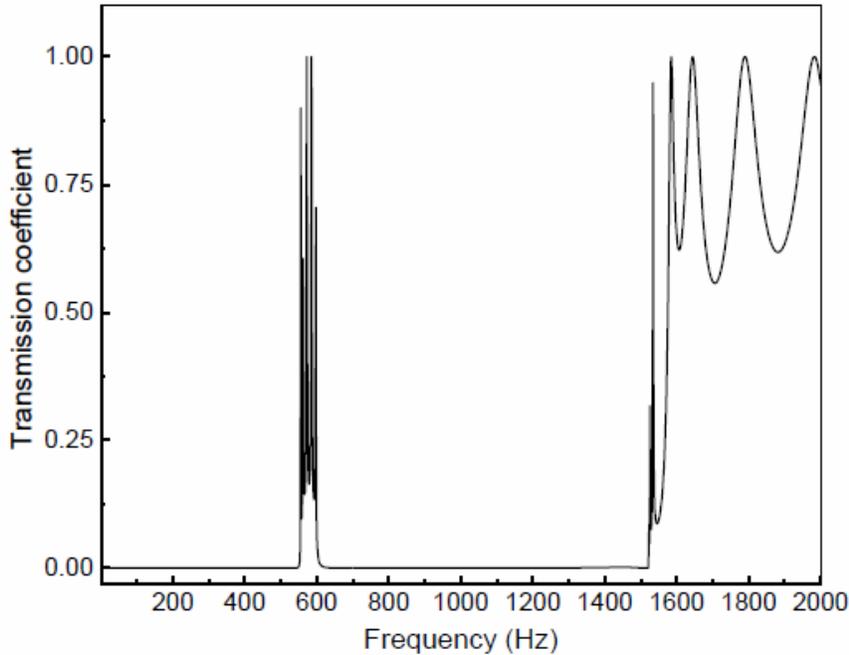

FIG. 10. Pressure transmission coefficient for a composite array of six unit cells.

Next, an array of 6 unit cells of the combined system was modeled and the pressure transmission characteristics were obtained. Note that arrays with more number of unit cells were also modeled, however, the only noticeable difference was in the number of local peaks and dips in the pass band which correspond to the number of unit cells used. Figure 10 shows the pressure transmission coefficient obtained. Two bands of zero transmission, 0Hz to 550 Hz and 600 Hz to 1520 Hz, separated by a passing band are clearly visible. The effective properties obtained using the inverse method show that the effective density and the effective modulus behavior of the combined systems is quite different from the individual systems, with the density and modulus being simultaneously negative only in the narrow bands of 550 Hz to 600 Hz and beyond 1520Hz to 2000 Hz, corresponding to the obtained pass band (Fig. 11). Thus, the properties of the composite structure are not the same as the individual structures.

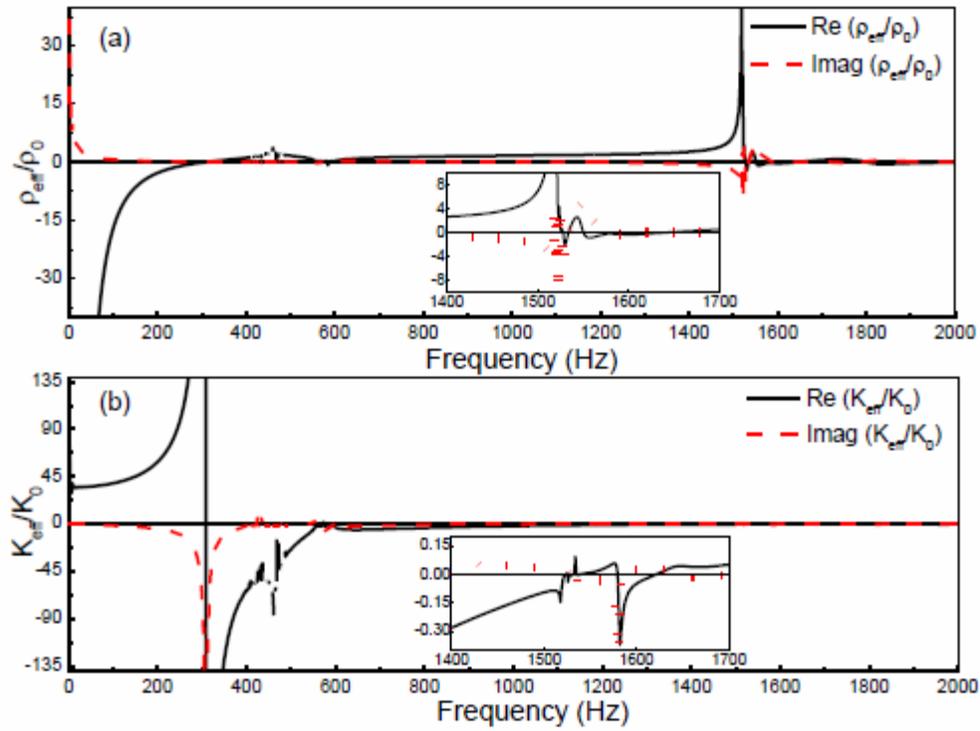

FIG. 11. (a) Variation of effective density and (b) effective modulus, of the composite structure. The insets show a magnified view of the two parameters from 1400 Hz to 1700 Hz.

## 5. Conclusions

A new design for a negative effective bulk modulus structure was proposed based on the traditional expansion chamber. A neck region was added to the expansion chamber to obtain high transmission loss behavior at low frequencies and it was shown that the structure displays negative effective bulk modulus at these frequencies. Additionally, a unified picture of membrane based metamaterials was presented. Both the existing designs were analyzed using FEM and the equivalence between the stretched membrane array design and the mass loaded membrane design was established. It was shown that both systems display negative effective density in two different frequency regions, in the lower frequency region due to the opposition offered by the prestress in the membrane and at a higher frequency region due to the antiresonance behavior. The modified expansion chamber design was combined with an array of stretched membranes to obtain a structure capable of displaying double negative behavior. It was shown that the effective properties of the combined system are different as compared to the properties of the individual systems, and only a narrow band of double negativity was obtained.

# Acknowledgement

This work was supported by an Office of Naval Research grant No. N00014-11-1-0580. Dr Yapa D.S. Rajapakse was the program manager.